\documentclass[aps,prl,twocolumn,superscriptaddress]{revtex4-1}

\usepackage{graphicx}
\usepackage[usenames,dvipsnames]{xcolor}
\usepackage[bookmarks=false,colorlinks]{hyperref}
\hypersetup{
	linkcolor=MidnightBlue,        
	citecolor=blue,     
	filecolor=Plum,      	
	urlcolor=MidnightBlue,           
}

\usepackage{ccaption}
\usepackage{ragged2e}
  \captionnamefont{\small \bfseries }
  \captiontitlefont{\small }
  \captionstyle{\justifying}

\begin{document}

\title{
Ultrafast quasiparticle dynamics in correlated semimetal Ca$_3$Ru$_2$O$_7$
}

\author{Yakun Yuan$^+$}
\affiliation{Materials Research Institute and Department of Materials Science $\&$ Engineering, Pennsylvania State University, University Park, Pennsylvania 16802, USA}
\author{Peter Kissin$^+$}
\affiliation{Department of Physics, University of California San Diego, La Jolla, California 92093, USA}
\author{Danilo Puggioni}
\affiliation{Department of Materials Science $\&$ Engineering, Northwestern University, Evanston, IL 60208, USA}
\author{Kevin Cremin}
\affiliation{Department of Physics, University of California San Diego, La Jolla, California 92093, USA}
\author{Shiming Lei}
\affiliation{Materials Research Institute and Department of Materials Science $\&$ Engineering, Pennsylvania State University, University Park, Pennsylvania 16802, USA}
\author{Yu Wang}
\affiliation{Department of Physics $\&$ Engineering Physics, Tulane University, New Orleans, Louisiana 70118, USA}
\author{Zhiqiang Mao}
\affiliation{Department of Physics $\&$ Engineering Physics, Tulane University, New Orleans, Louisiana 70118, USA}
\affiliation{Department of Physics, Pennsylvania State University, University Park, Pennsylvania 16802, USA}
\author{James M. Rondinelli}
\affiliation{Department of Materials Science $\&$ Engineering, Northwestern University, Evanston, IL 60208, USA}
\author{Richard D. Averitt}
\email[E-mail:]{raveritt@ucsd.edu}
\affiliation{Department of Physics, University of California San Diego, La Jolla, California 92093, USA}
\author{Venkatraman Gopalan}
\email[E-mail:]{vxg8@psu.edu}
\affiliation{Materials Research Institute and Department of Materials Science $\&$ Engineering, Pennsylvania State University, University Park, Pennsylvania 16802, USA}
\affiliation{Department of Physics, Pennsylvania State University, University Park, Pennsylvania 16802, USA}

\date{\today}

\begin{abstract}
The correlated polar semimetal Ca$_3$Ru$_2$O$_7$ exhibits a rich phase diagram including two magnetic transitions ($T_N$=56 K and $T_C$=48 K) with the appearance of an insulating-like pseudogap (at $T_C$). In addition, there is a crossover back to metallic behavior at $T^*$=30 K, the origin of which is still under debate. We utilized ultrafast optical pump optical probe spectroscopy to investigate quasiparticle dynamics as a function of temperature in this enigmatic quantum material. We identify two dynamical processes, both of which are influenced by the onset of the pseudogap. This includes electron-phonon relaxation and, below $T_C$, the onset of a phonon bottleneck hindering the relaxation of quasiparticles across the pseudogap. We introduce a gap-modified two-temperature model to describe the temperature dependence of electron-phonon thermalization, and use the Rothwarf-Taylor to model the phonon bottleneck. In conjunction with density functional theory, our experimental results synergistically reveal the origin of the $T$-dependent pseudogap. Further, our data and analysis indicate that $T^*$ emerges as a natural consequence of $T$-dependent gapping out of carriers, and does not correspond to a separate electronic transition. Our results highlight the value of low fluence ultrafast optics as a sensitive probe of low energy electronic structure, thermodynamic parameters, and transport properties of Ruddlesden-Popper ruthenates.

\end{abstract}

\maketitle

\begin{figure}
 \includegraphics[width=1\columnwidth]{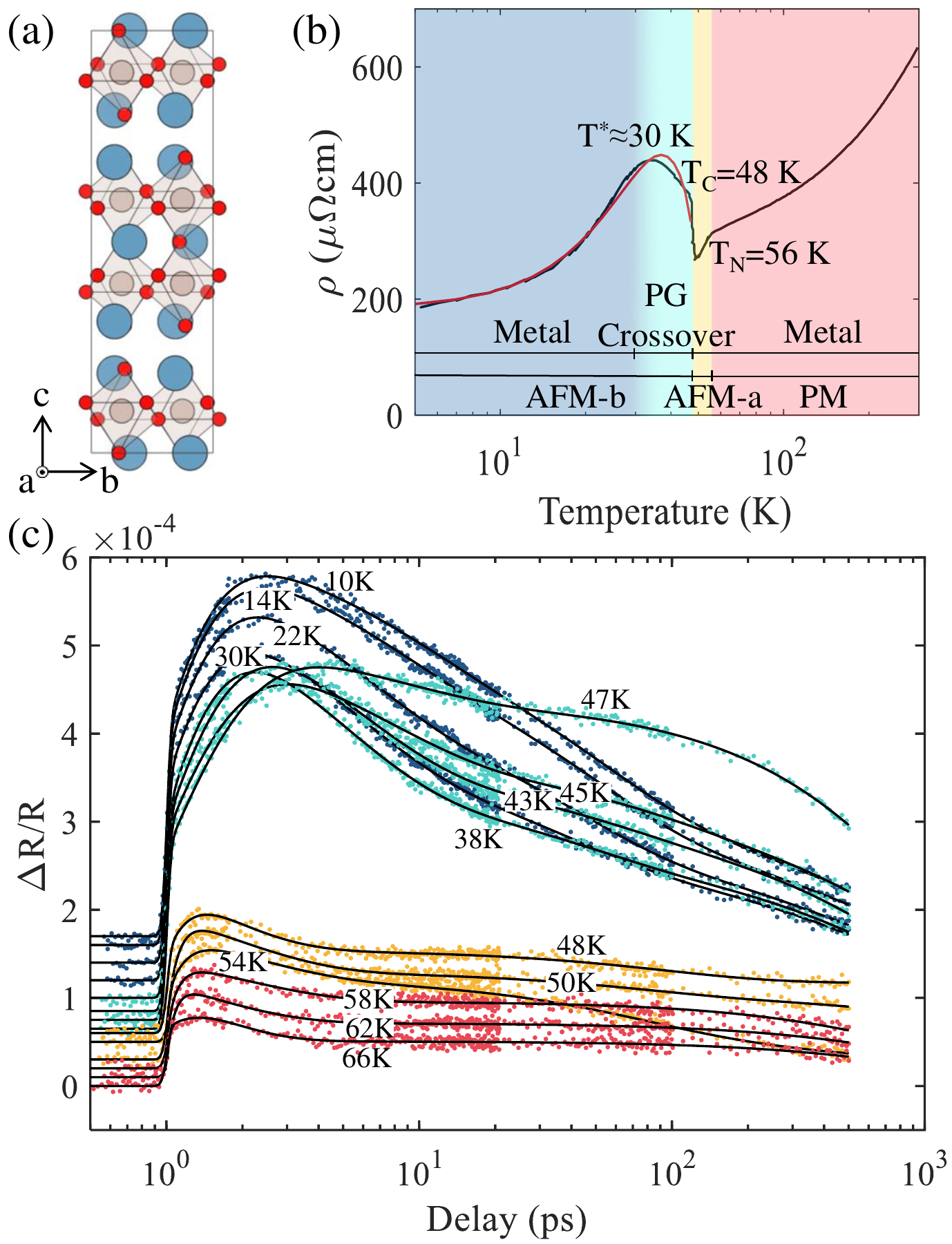}
 \caption{(a) Atomic structure of Ca$_3$Ru$_2$O$_7$ with a $Bb2_1m$ space group. Ca, Ru, and O atoms are labeled in blue, brown and red, respectively. (b) DC resistivity of  Ca$_3$Ru$_2$O$_7$ in the a-b plane\cite{Lee2007}. Regions with different colors represent different behaviors emerging at low temperature. PM stands for paramagnetic phase. The red line is a fit based on $T$-dependent carrier concentration and scattering rate discussed later. (c) Temperature dependent OPOP data on Ca$_3$Ru$_2$O$_7$. Four distinct relaxation behaviors are clearly observed at various temperatures, which are color-coded to match the resistivity phase diagram in (b). Black lines are fits using the multi-exponential decay model.
 \label{OPOP}}
\end{figure}

Ruddlesden-Popper ruthenates have received tremendous research interest since the discovery of superconductivity in Sr$_2$RuO$_4$, which is the only non-copper-based superconductor isostructural to La$_{2-x}$(Sr,Ba)$_x$CuO$_4$ \cite{Mackenzie1998,Maeno1994}. Previous studies on Ca$_3$Ru$_2$O$_7$ (space group $Bb2_1m$, Fig. \ref{OPOP}(a)) have revealed a rich interplay between spin, lattice and electronic degrees of freedom \cite{Liu1999,Lee2007,Cao1997,Peng2013,Yoshida2004,Lei2018}. As shown in Fig. \ref{OPOP}b, between $T_C$=48 K and $T_N$=56 K, the spins in Ca$_3$Ru$_2$O$_7$ align ferromagnetically within the a-b plane and antiferromagnetically (AFM) along c-axis, with the spins oriented along the a-axis (AFM-a). At $T_C$=48 K, the resistivity begins to display insulating behavior and the spins align along the b-axis (AFM-b) \cite{Peng2013}, with metallic resistivity reemerging below $T^*$=30 K \cite{Yoshida2004,Cao1997,Lee2007}.

The electronic structure changes of Ca$_3$Ru$_2$O$_7$ at $T_C$ and $T^*$ require further clarification. $T_C$ was previously thought to correspond to a Mott-like transition \cite{Cao1997,Cao2004}, however, the observation of a partial gap opening near E$_F$ points to a Fermi surface (FS) instability \cite{Baumberger2006}, which suggests the appearance of a density wave \cite{Lee2007, Kikugawa2010}, despite the unusual redistribution of optical spectral weight. While a density wave has not been directly observed in Ca$_3$Ru$_2$O$_7$, this scenario may offer new opportunities to study the emergence of charge and spin density waves in the isostructural bilayer cuprates (see Fig. \ref{OPOP}(a)). Recently, a Lifshitz transition was proposed as the origin of the resistivity crossover at $T^*$ \cite{Xing2018}. 

The interactions between strongly correlated degrees of freedom are often more easily disentangled when studied on their natural timescales, motivating dynamical studies of Ca$_3$Ru$_2$O$_7$ in the time domain. Optical pump optical probe (OPOP) spectroscopy has been used to study the relaxation dynamics of photoexcited quasiparticles (QPs) in a variety of materials \cite{Kabanov1999, Demsar1999, Demsar2003, Demsar2006, Ghosh2017, Park2018, Batignani2018}. The versatility of this technique derives from its extreme sensitivity to the formation of small gaps in the electronic density of states (DOS) near the Fermi energy E$_F$. The presence of a gap can be inferred from the temperature and pump fluence dependence of the QP relaxation dynamics, and may result in an increase of the relaxation time by several orders of magnitude. OPOP complements conventional frequency domain techniques in characterizing the low energy electronic structure of quantum materials.

In this letter, we investigate quasiparticle dynamics in Ca$_3$Ru$_2$O$_7$ using OPOP spectroscopy.  Our data reveals the development of a pseudogap that dramatically alters the relaxation dynamics, slowing down the electron-phonon relaxation near Fermi surface, giving rise to a new relaxation component related to a phonon bottleneck associated with above-gap QP excitations. The coexistence of both relaxation channels with comparable strength in one material is unexpectedly rare, making Ca$_3$Ru$_2$O$_7$ the first platform to study quasiparticle dynamics affected by both pseudogap and Fermi surface. Analysis with a $T$-dependent DOS  shows that the crossover at $T^*$ can be explained as a consequence of the $T$-dependent gap that opens at $T_C$, without invoking a separate electronic transition, simplifying the understanding of the phase diagram of Ca$_3$Ru$_2$O$_7$.

OPOP measurements on Ca$_3$Ru$_2$O$_7$ used 25 fs laser pulses centered at 800 nm with a repetition rate of 209 kHz. The pump fluence was fixed at 2 $\mu$J/cm$^{2}$ for all reported measurements to ensure minimal sample heating. The sample temperature change is determined to be less than 1 K under experimental temperature $>$20 K, which increases to 7 K at the lowest experimental temperature 10 K (see analysis in Supplementary Sec. I and II \cite{Sup}). The probe fluence was 1 $\mu$J/cm$^{2}$. Due to the low repetition rate, the system was able to fully relax back to ground state before each pump pulse. And the probe fluence dependent test verified there was no probe induced effect. The cross polarized pump and probe beams were focused on to the sample to $1/e^2$ spot diameters of 70 $\mu$m and 40 $\mu$m respectively. The data were collected from large, flat areas of samples cleaved in the a-b plane. A continuous flow liquid Helium optical cryostat was used for temperature control.

Figure \ref{OPOP}(c) shows the $T$-dependence of the photoinduced change in fractional reflectivity \(\Delta\)R/R as a function of time from 66 K to 10 K. The data are plotted in logarithmic x-scale with time zero shifted to 1 ps for better visualization. The temporal resolution is 25 fs. Above 66 K, the dynamics are only weakly $T$-dependent (For additional data, see Supplementary Sec. I \cite{Sup}). Above $T_C$, the relaxation is biexponential, consisting of a large, fast component that relaxes on a subpicosecond timescale and a small, slow component that persists beyond the measurement window of 500 ps. Near $T_N$, between 58 K and 54 K, the fast component slows and the slow component becomes faster, marking the PM to AFM-a transition. At \(T_{C}\), between 48 K and 47 K, the maximum signal amplitude increases by a factor of three and the relaxation dynamics become triexponential, with a new relaxation process emerging on an intermediate timescale of nearly 100 ps. These abrupt changes to the dynamics mark the first order transition to the partially gapped low temperature phase\cite{Cao1997}. Cooling from 47 K to 10 K, the fast component first becomes faster then slows down, showing a minimum signal amplitude at 10 ps near $T^*$.

We fit the data to a multiexponential function of the form \cite{Demsar2006}:
\begin{equation}
\frac{\Delta R}{R}(t) = f(t)\times(A_{1}e^{-t/\tau_{d1}}+A_{2}e^{-t/\tau_{d2}}+A_{3}e^{-t/\tau_{d3}})\label{eq1}
\end{equation}
Where $f(t)=r\times(\frac{1}{2}+\frac{1}{2}erf(\sqrt{2}(t-t_{0})/\tau_{p}))+(1-r)\times(1-e^{-(t-t_{0})/\tau_{r}}))$. In $f(t)$, the first term containing the error function represents the cross-correlation of the pump and probe pulses with a $T$-independent pulse duration, $\tau_p$. The second term containing $\tau_{r}$ represents the slow rise dynamics that onset mainly below $T_C$. $r$ and $1-r$ are weights for above two contributions, respectively. The main term, which contains three strongly $T$-dependent exponential decays, is the focus of our analysis. The fast component $A_1,\tau_{d1}$ and the new component $A_2,\tau_{d2}$ emerging below $T_C$ are associated with QP dynamics near E$_F$. The slow component $A_3,\tau_{d3}$ arises from the thermal dissipation of pump induced energy. (see Supplementary Sec. I for all fitting parameters and Sec. II for heat capacity and thermal conductivity extracted from  $A_3$ and $\tau_{d3}$)\cite{Sup}.

\begin{figure}
 \includegraphics[width=1\columnwidth]{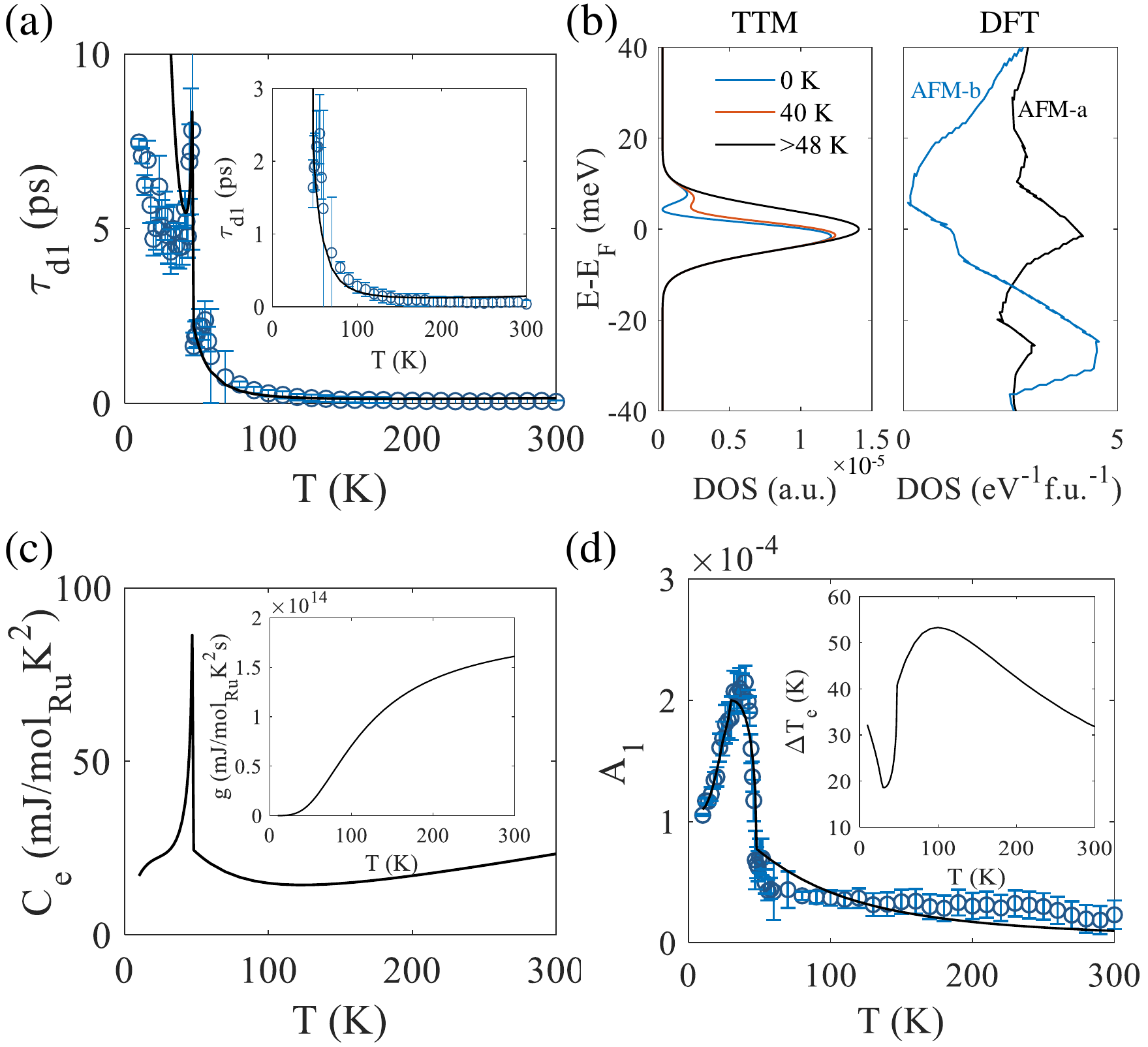}
 \caption{(a) Temperature dependence of the electron-phonon thermalization time constant $\tau_{d1}$. (b) Temperature dependence of DOS near E$_F$ obtained from TTM analysis (left panel). DOS for AFM-a and AFM-b ground states predicted by DFT. (right panel) (c) Electronic heat capacity heat from TTM calculation. The inset shows the temperature dependent electron-phonon coupling constant near the Fermi surface used for TTM analysis. (d) Amplitude of electron-phonon thermalization dynamics $A_1$ as a function of temperature. The fit takes values in from the DOS in (b) and the electronic specific heat yielded from (a) and shown in (c). The inset shows the electronic temperature change upon optical pump.
 \label{FastDyn}}
\end{figure}

We first focus on the fast relaxation component $A_1$ and $\tau_{d1}$. As shown in Fig. \ref{FastDyn}(a), the time constant of this fast component is sub-picosecond at high temperatures and slows down upon approaching $T_C$. Within this temperature range, Ca$_3$Ru$_2$O$_7$ exhibits metallic behavior ($\frac{d\rho}{dT}>0$) as shown by Fig. \ref{OPOP}(b). Following photoexcitation, the fast relaxation dynamics of metals are governed by electron-phonon (e-ph) thermalization, which can be described using the two-temperature model (TTM)\cite{Groeneveld1995,Demsar2006}. A central assumption of the TTM is electron-electron (e-e) scattering is much faster than electron-phonon (e-ph) scattering. Thus, the photoexcited electron subsystem can almost immediately be described with an elevated electronic temperature $T_e$. Subsequently, e-ph thermalization increases the lattice temperature ($T_l$), until a quasiequilibrium is reached between the two subsystems.

The e-ph thermalization time constant in the TTM is given by $\tau_{e-ph}=\frac{1}{g}\frac{C_eC_l}{C_e+C_l}$, where $g(T)$ is the e-ph coupling function, $C_e$ and $C_l$ are the specific heats for the electron and lattice subsystems, respectively. Typically, $C_l \gg C_e$ at high temperatures. On approaching 0 K, $C_e=\gamma T$ and $C_l=\beta T^3$ , with $\gamma$=1.7 mJ/mol$_{Ru}$K$^2$ and $\beta$=0.14 mJ/mol$_{Ru}$K$^4$ for Ca$_3$Ru$_2$O$_7$\cite{Yoshida2004}. At the lowest temperature (10 K) of this OPOP study, $C_e$=17$\ll$140 mJ/mol$_{Ru}$K=$C_l$. Thus, the e-ph thermalization time at all temperatures can be further simplified as:
\begin{equation}
 \tau_{e-ph}\approx\frac{C_e}{g}
 \label{eq4}
 \end{equation}
We approximate $g(T)$ for Ca$_3$Ru$_2$O$_7$ using an expression valid for simple metals $g(T)=dG(T)/dT, G(T)=4g_\infty(T/\theta_D)^5\int_0^{\theta_D/T}\frac{x^4}{e^x-1}dx$ \cite{Demsar2006}. The Debye temperature $\theta_D$=437 K (For calculations, see Supplementary Sec. III\cite{Sup}), leaving the constant $g_\infty$ as the only free parameter. The $T$-dependence of $g(T)$ is plotted in the inset of Fig. \ref{FastDyn}(c). Assuming a $T$-independent $C_e$ in this temperature range, a fit to $\tau_{d1}$ above $T_C$, shown in the inset of Fig. \ref{FastDyn}(a), agrees well with our data, supporting the TTM description of the dynamics above $T_C$.

To account for changes in e-ph thermalization dynamics in the vicinity of Tc (e.g. both the lifetime and amplitude) we study the effect of a gap on the e-ph relaxation time by considering a phenomenological $T$-dependent DOS of the form:
\begin{equation}
D_e(\epsilon)=D_0+D_1e^{-\frac{\epsilon^2}{w_1^2}}[1-\frac{\Delta(T)}{\Delta_0}e^{-\frac{(\epsilon-\epsilon')^2}{w_2^2}}] \label{eq5}
\end{equation}
The summation of a constant, $D_0$, and a Gaussian with width $w_1$ is often used to describe the DOS of correlated metals.\cite{Demsar2006} The term in the square brackets describes the reduction in the DOS due to the opening of a gap, with a BCS-like temperature dependent behavior, $\Delta(T)=\Delta_0tanh2.2\sqrt{\frac{T_C}{T}-1}$, where $\Delta(T)$ is the temperature dependent gap size, and $\Delta_0$ is the gap size at 0 K. The Gaussian centered at $\epsilon'$ with width $w_2$ approximates the shape of the DOS reduction due to the gap opening. The electronic specific heat can then be derived from the electronic DOS near E$_F$ by $C_e=\frac{\partial}{\partial T}\int_{-\infty}^\infty \epsilon D_e(\epsilon,T)f(\epsilon,T)d\epsilon, \epsilon=E-E_F$, ignoring the weak temperature dependence of E$_F$.

Allowing the parameters of the model DOS Eq.\ref{eq5} to vary, we fit $\tau_{d1}$ using Eq.\ref{eq4} for all temperatures. The fit, plotted in Fig. \ref{FastDyn}(a), reproduces the behavior down to $\sim$40 K, including the slow down as cooling down to $T_C$, the sudden jump at $T_C$ and the subsequent drop below $T_C$. The deviation of the fit below $\sim$40 K signifies the failure of the TTM at very low temperatures, which is quite general in TTM analysis because the timescale of e-e thermalization slows down and becomes comparable to the timescale of e-ph thermalization at low $T$ \cite{Demsar2006}. The extracted DOS described by Eq.\ref{eq5} and the resulting electronic specific heat $C_e$ are plotted in Figs. \ref{FastDyn}(b) left panel and (c) respectively, with parameters of $w_1$=6.3$\pm$0.3 meV, $\epsilon'$=4$\pm$2 meV, $w_2$=4$\pm$2 meV. The gap in DOS centered at 4$\pm$2 meV qualitatively agrees with 8 meV gap observed by ARPES \cite{Baumberger2006}. Furthermore, key features of our model DOS, including a gap just above E$_F$ and a large peak just below E$_F$, are reproduced by density functional theory with static correlations (DFT+$U$+$SOC$), as shown in the right panel of Fig. \ref{FastDyn}(b), supporting the DOS extracted from TTM analysis (See Supplementary Sec. IV\cite{Sup} for details). In addition, we would like to point out that a gap below the Fermi level fails to fit the experimental data.

In addition, the transient amplitude $A_1$ can be modeled with the TTM by considering the number density of thermally activated electrons \cite{Demsar2006}:
\begin{equation}
A_1\propto n_{T_e'}-n_{T_e}=\int_0^\infty D_e(\epsilon,T_e')f(\epsilon,T_e')-D_e(\epsilon,T_e)f(\epsilon,T_e)d\epsilon    \label{eq6}
\end{equation}
The electronic temperature after photoexcitation $T_e'$ is determined by $\Delta U=\int_{T_e}^{T_e'}C_edT$, with deposited energy density $\Delta U$. Equation (\ref{eq6}) considers both changes in DOS and carrier population. The transient electronic temperature change $\Delta T_e$, plotted in the inset of Fig. \ref{FastDyn}(d), is substantial, thus a transient suppression of the gap and a corresponding change in $D_e(\epsilon,T)$ at $T_e$ and $T_e'$ in Eq.\ref{eq6} must be considered in order to reproduce the $T$-dependence of $A_1$. Taking $D_e(\epsilon,T)$ from Fig. \ref{FastDyn}(b), the fit to $A_1$ using Eq.\ref{eq6} is shown in Fig. \ref{FastDyn}(d). The excellent fit captures all characteristics over the entire temperature range of the measurement, which strongly suggests that the TTM model captures essential features of the role of the pseudogap in determining the e-ph thermalization. In addition, we also point out that the phonon bottleneck picture described later by the Rothwarf-Taylor model fails to explain this dynamics (see Supplementary Sec. V\cite{Sup}).

The development of the gap is expected to deplete the free carriers in the system, resulting in an increase of resistivity below $T_C$. On the other hand, the decrease in carrier scattering rate at low temperatures will reduce the resistivity. Using the DOS employed for the TTM modeling, we investigate if the $T$-dependent carrier density and scattering time is responsible for the upturn in resistivity ($\frac{d\rho}{dT}< 0$) between $T^*$ and $T_C$. We approximate the population of free electrons ($n$), and holes ($p$), to be: $n=\int_{-\infty}^\infty D_e(\epsilon,T)f(\epsilon,T)d\epsilon-N_1$ and $p=\int_{-\infty}^\infty D_e(\epsilon,T)[1-f(\epsilon,T)]d\epsilon-N_2$, where $N_1$ and $N_2$ are populations of localized electrons and holes near E$_F$. The total free carriers can be written as $n+p=\int_{-\infty}^\infty D_e(\epsilon,T)d\epsilon-N, N=N_1+N_2$. The scattering rate follows Fermi liquid behavior with impurities at low temperatures: $1/\tau\propto T^2+\gamma_0$. Thus, the resistivity is expressed as: $\rho\propto\frac{1}{\tau(n+p)}$. The fits to experimental data is plotted as red line in Fig. \ref{OPOP}(b) (see Supplementary Sec. VI\cite{Sup} for $n+p$ and $1/\tau$). The great quality of the fit suggests the insulating behavior between $T^*$ and $T_C$ arises from the depletion of free carriers by the opening of the pseudogap, and that the insulating to metallic resistivity crossover at $T^*$ is a crossover temperature that originates from the competition between carrier scattering rate and population rather than from a separate electronic transition.

\begin{figure}
 \includegraphics[width=1\columnwidth]{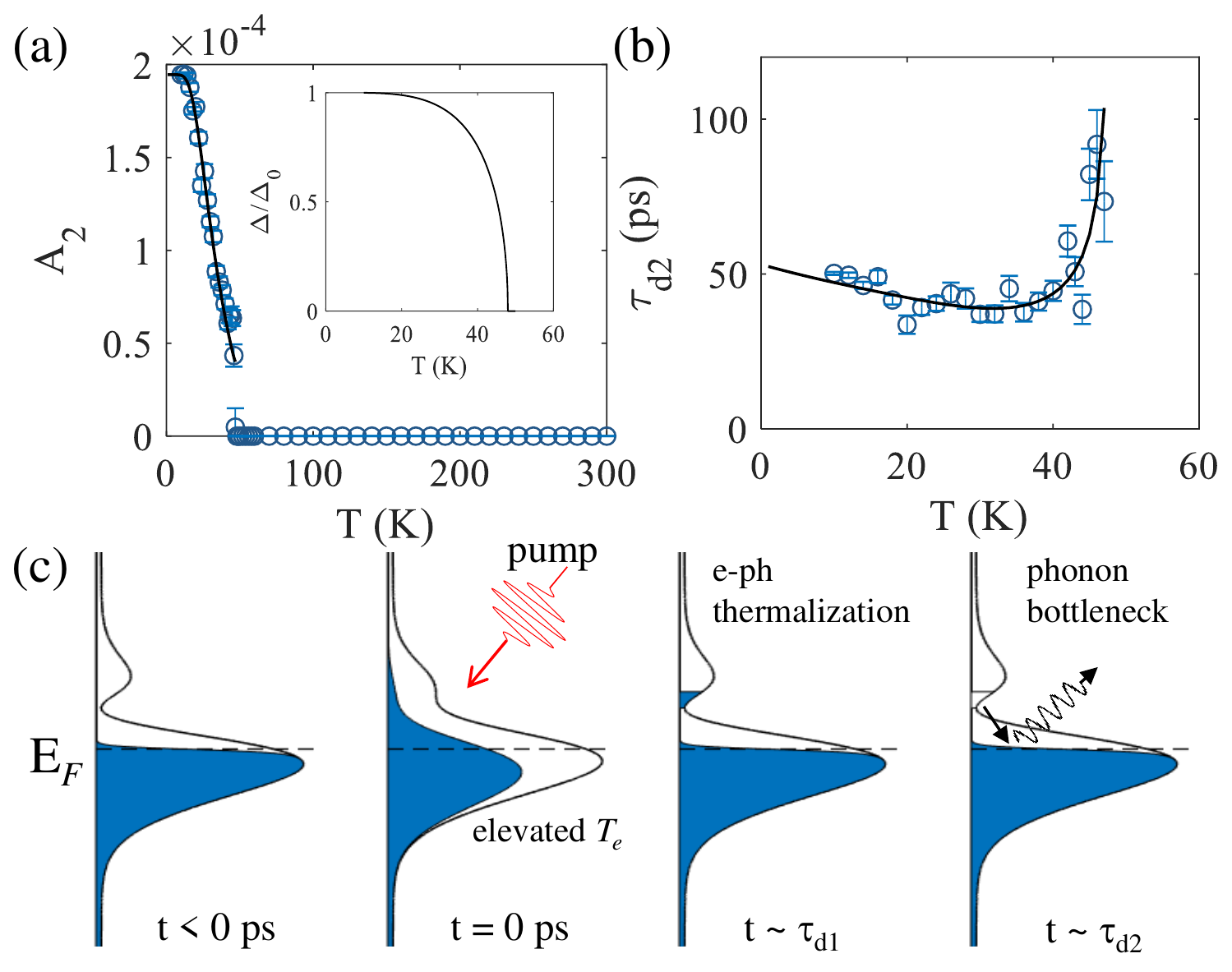}
 \caption{(a) Amplitude $A_2$ of QP dynamics emergent below $T_C$=48 K. The R-T fit yields a BCS-like gap with gap size of 7.1$\pm$0.2 meV at 0 K. The inset shows the $T$-dependence of the gap size $\Delta/\Delta_0$.(b) Time constant $\tau_{d2}$ of QP relaxation dynamics. The R-T fit reveals a consistent gap size of $\Delta_0$=7.1$\pm$0.2 meV and $\alpha=2.39\times10^{-8}<\frac{52}{\theta_D^3T_{min}}=6.23\times10^{-8}$. (c) Schematics of photoexcited QPs relaxation dynamics at different stages.
 \label{NewDyn}}
\end{figure}

We now turn to the new relaxation process $A_2$ and $\tau_{d2}$ that emerges below $T_C$. In a diverse assortment of other systems with small gaps near E$_F$, the timescale of the return to equilibrium is determined by the $meV$-scale kinetics of electron-hole (e-h) recombination. These kinetics may be described with the phenomenological Rothwarf-Taylor (R-T) model \cite{Rothwarf1967} (For a detailed discussion of the R-T model see Supplementary Sec. VII\cite{Sup}).

In the small photo excitation limit \cite{Kabanov2005}, the R-T model relates the density of thermally activated QPs, $n_T$, to the measured transient reflectivity amplitudes, $A$, and relaxation rates, ${\tau}^{-1}$, as \cite{Demsar2006,Kabanov2005,Chia2006}:

\begin{equation}
n_T(T)\propto\frac{A(T\to0)}{A(T)}-1\label{eq2}
\end{equation}
\begin{equation}
\tau^{-1}(T) = \Gamma[\delta+2n_{T}(T)](\Delta(T)+\alpha \Delta(T)T^{4})\label{eq3}
\end{equation}

where $\Gamma$, $\delta$, and $\alpha$ are $T$-independent fitting parameters, $\Delta(T)+\alpha \Delta(T)T^{4}$ describes the dependence of the high energy phonon (HEP) decay rate on the gap size $\Delta$, with an upper limit of $\alpha<52/(\theta^3_DT_{min})$\cite{Chia2006}, where $T_{min}$=10 K is the minimum temperature of the experiment. For our analysis, we assume a standard form of the thermal QP density $n_T(T)\propto\sqrt{\Delta(T)T}exp(-\Delta(T)/T)$ \cite{Kabanov1999,Chia2006,Chia2007} and a BCS-like gap of the form $\Delta(T)=\Delta_0tanh2.2\sqrt{\frac{T_C}{T}-1}$.

The amplitude $A_2$, and thus the thermal QP density implied by Eq.\ref{eq2}, can be fitted as shown in Fig. \ref{NewDyn}(a), yielding a gap size of $\Delta_0$=7.1$\pm$0.2 meV=1.71$\pm$0.05$k_BT_C$, which is close to the BCS value of 1.76. Additionally, the abrupt onset of slow rise dynamics (For rise time, see supplementary Sec. VII\cite{Sup}) are most easily explained in the strong bottleneck regime of the R-T model. Thus, we conclude that the new relaxation process arises from a phonon bottleneck due to the presence of a gap in the DOS near E$_F$. Note that the 7.1$\pm$0.2 meV gap size is in excellent agreement with 8 meV band edge shift revealed by ARPES\cite{Baumberger2006}. However, it is smaller than the 13 meV reported by optical spectroscopy measurement\cite{Lee2007}, suggesting this is an indirect gap. Using the above extracted gap size $\Delta(T)$, the relaxation time $\tau_{d2}$ is fitted by Eq.\ref{eq3} as shown in Fig. \ref{NewDyn}(b). The quality of the fits confirms the presence of a BCS-like indirect gap below $T_C$ and the associated phonon bottleneck picture.

With above analysis, we construct a comprehensive picture of photoexcited QPs relaxation dynamics in Ca$_3$Ru$_2$O$_7$ as schematically shown in Fig. \ref{NewDyn}(c). Below $T_C$, optical excitation at $t$=0 ps increases the electronic temperature, $T_e$, leading to a wider range of occupancy near E$_F$ and results in a transient suppression of the pseudogap. The subsequent e-ph thermalization decreases the electronic temperature together with a rapid recovery of the gap on a timescale of $\sim\tau_{d1}$. However, the electrons above the gap relax on a different timescale $\sim\tau_{d2}$ limited by the kinetics of the phonon bottleneck. The final slow recovery ($\sim \tau_{d3}$) is governed by thermal transport into the bulk of the sample, from which the $T$-dependent heat capacity and thermal conductivity are extracted (see Supplementary Sec. II\cite{Sup}).

In summary, we present the first ultrafast optical spectroscopic study of the pseudogap phase of Ca$_3$Ru$_2$O$_7$. The $T$-dependent DOS yielded from the fast e-ph thermalization near Fermi surface together with the phonon bottleneck emerging below $T_C$ provide a synergistic picture of the indirect gap. From this picture, we are able to explain the insulating to metallic crossover in resistivity at $T^*$ as a natural consequence of the $T$-dependent emergence of the gap, without invoking a separate electronic transition. Given the simplicity of the TTM analysis described above, the consistency between the results yielded by the TTM on the fast dynamics and R-T model on the new dynamics below $T_C$ is surprisingly good. As demonstrated by this work, the incorporation of a $T$-dependent DOS into the TTM can capture the influence of a gap near E$_F$, providing an alternative approach to unveil the evolution of low energy electronic structure of strongly correlated metals in the time domain.

Y.Y., P.K., K.C., S.L., R.D.A. and V.G. were supported by the U.S. Department of Energy (U.S. DOE), Office of Basic Energy Sciences (BES) under Grant No. DE-SC00012375 for the experimental work, data analysis and the travel. D.P and J.M.R. were supported by the Army Research Office under Grant No. W911NF-15-1-0017 for the DFT studies. Y.W. and Z.M. was supported by the U.S. DOE under EPSCoR Grant No. DE-SC0012432 with additional support from the Louisiana Board of Regents for the crystal growth and characterization.\\
$^+$Y.Y. and P.K. contributed equally to this work.


\section{
Supplementary Materials: Ultrafast quasiparticle dynamics in correlated semimetal Ca$_3$Ru$_2$O$_7$
}

\section{I. Optical-pump optical probe data at 60-300K and the parameters of relaxation dynamics }
\begin{figure}[h]
 \includegraphics[width=0.8\columnwidth]{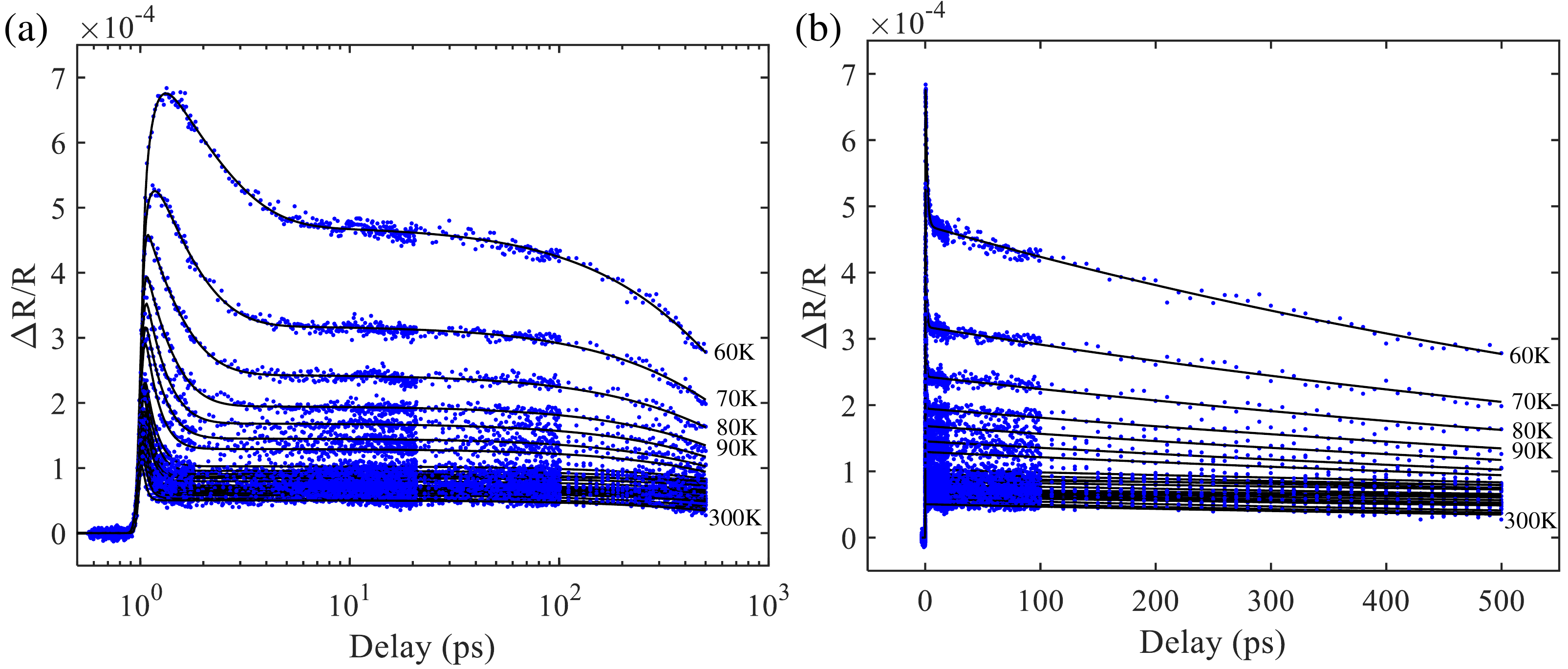}
 \caption{(a) Logarithmic and (b) Linear x-scale plots for $T$-dependent photo-induced reflectivity measurements on Ca$_3$Ru$_2$O$_7$ at high temperature. The time zero in (a) is shifted to 1 ps to avoid divergence at log(0).} A pump fluence of 10 $\mu$J/cm$^2$ was used. Black lines are fits using biexponential decay.
 \label{SHighOPOP}
\end{figure}

\begin{figure}[h]
 \includegraphics[width=0.8\columnwidth]{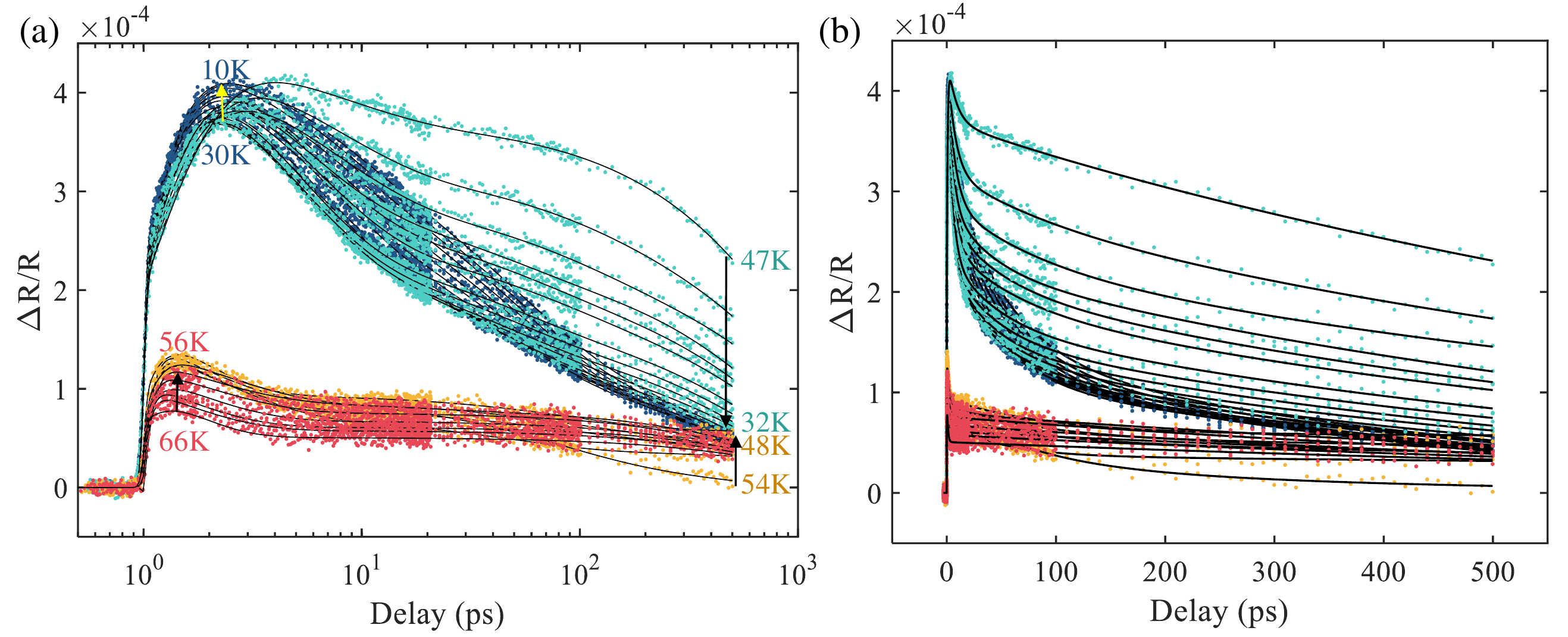}
 \caption{(a) Logarithmic and (b) Linear x-scale plots for $T$-dependent photo-induced reflectivity measurements on Ca$_3$Ru$_2$O$_7$ at low temperature. The time zero in (a) is shifted to 1 ps to avoid divergence at log(0).} A pump fluence of 2 $\mu$J/cm$^2$ was used. Black lines are fits using multiexponential decay.
 \label{SLowOPOP}
\end{figure}

\begin{figure}[h]
 \includegraphics[width=1\columnwidth]{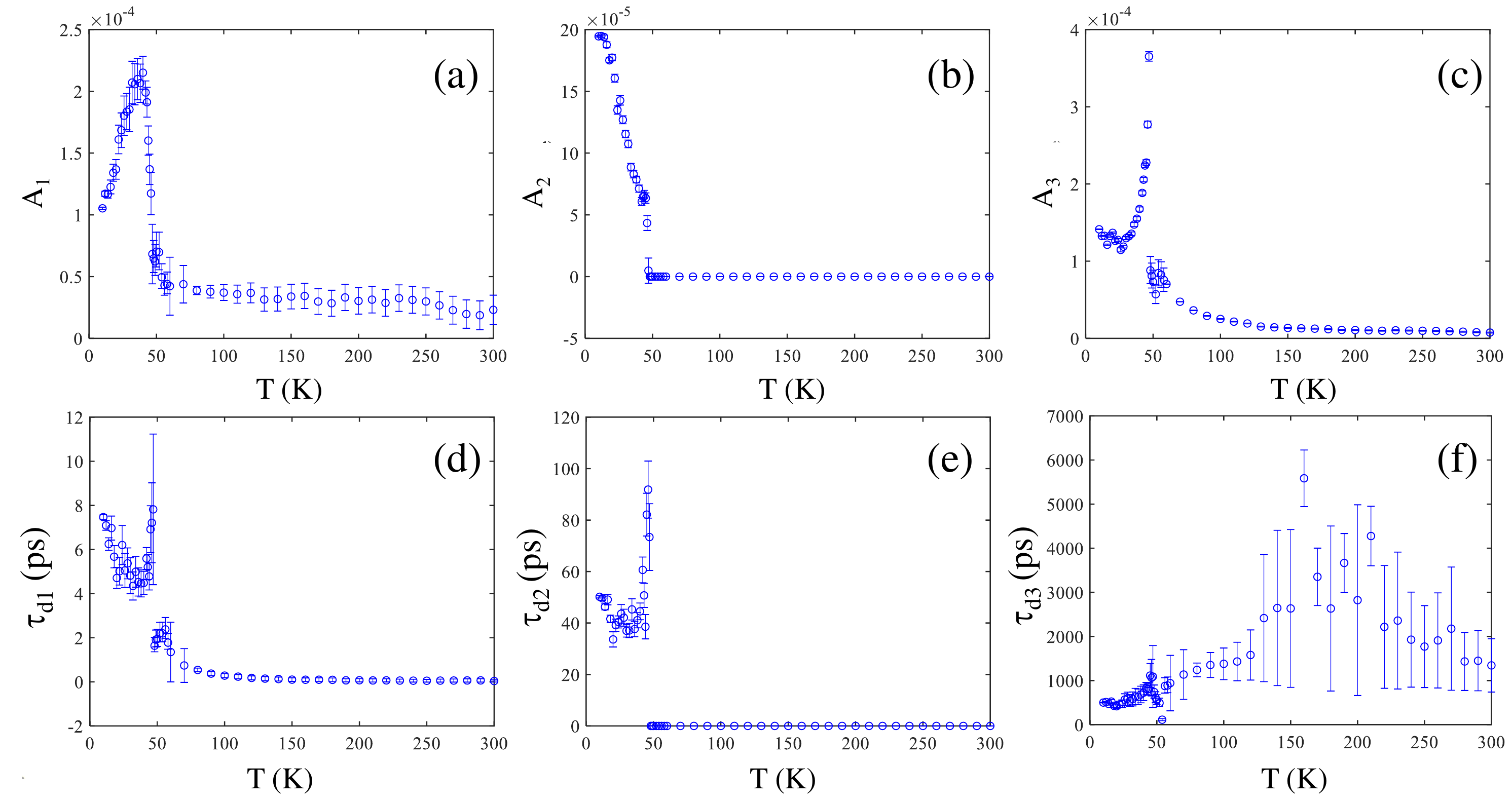}
 \caption{Amplitudes and time constants of the three experimentally observed relaxation dynamics. (a) and (d) are for the fast dynamics present at all temperatures. (b) and (e) are for the new dynamics appearing below $T_C$. (c) and (f) are for the slow dynamics due to thermal dissipation process. 
 \label{SPara}}
\end{figure}

\section{II. Slow dynamics and thermal properties of $Ca_3Ru_2O_7$}
\begin{figure}[h]
 \includegraphics[width=1\columnwidth]{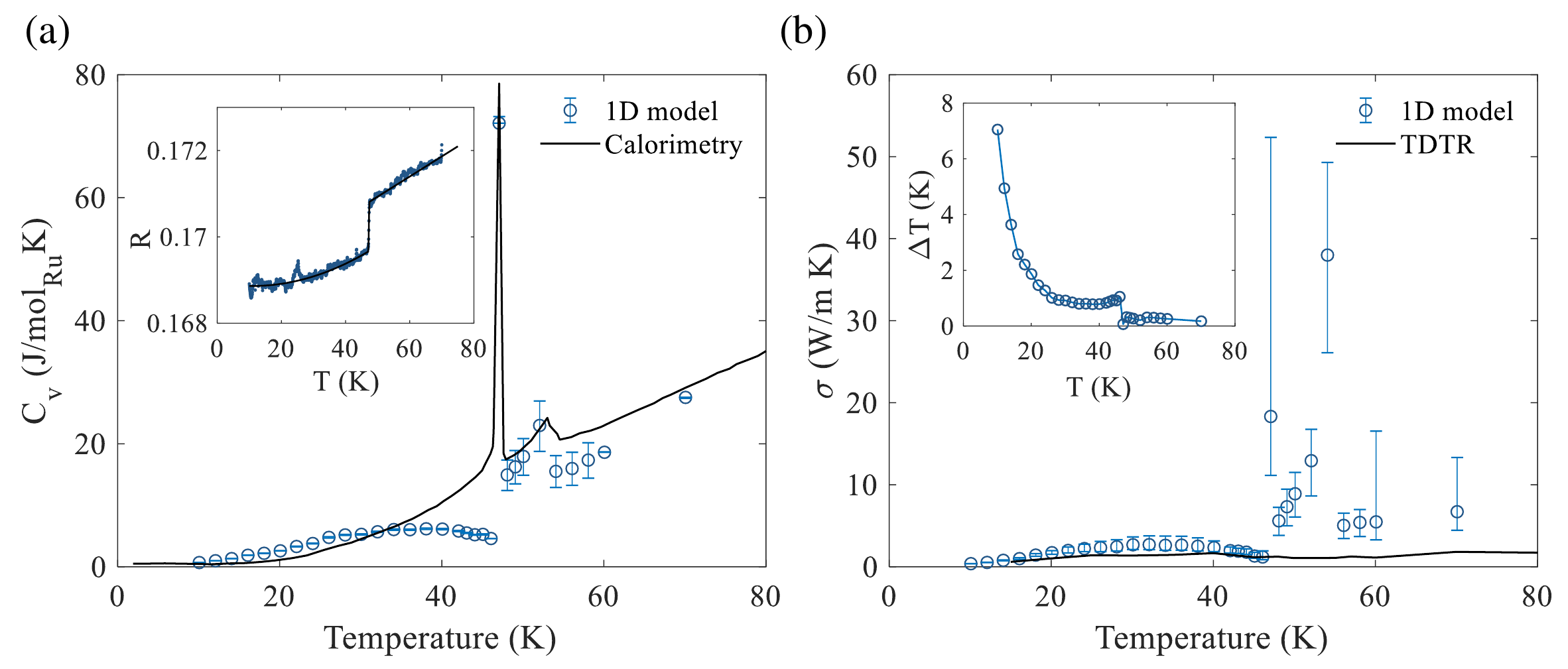}
 \caption{(a) Heat capacity calculated by modeling the amplitude of the slow dynamics $A_3$ with the  one-dimensional (1D) heat transport model described in the text. The black line is obtained by calorimetry measurement from literature.\cite{Ke2014}. The inset is the temperature dependent reflectivity of Ca$_3$Ru$_2$O$_7$ at 800 nm used for the calculation. (b) The thermal conductivity extracted by using the time constant of the slow dynamics $\tau_{d3}$ under 1D model (open circles) agrees qualitatively with TDTR results (black line). The inset is the maximum temperature change induced by optical pump after electron-phonon thermalization given by 1D model.
 \label{SSlowDyn}}
\end{figure}

The slow relaxation dynamics, $A_3$ and $\tau_{d3}$ (see Fig. \ref{SPara}(c) and (f)), are analyzed as follows. After e-ph thermalization and decay of HEP during the phonon bottleneck process results in a quasiequilibrium between electron and lattice subsystems, thermal conductivity (out of the probe volume) occurs with time scale of $>$500 ps, which we model qualitatively. Assuming that the heat transport only occurs perpendicular to the sample surface, and that the sample instantly establishes an exponential temperature profile upon optical pumping with characteristic length same as the optical penetration depth, an one-dimensional (1D) solution to the surface temperature evolution can be written as:
\begin{equation}
\Delta T=\frac{(1-R)4\pi kI_0}{\lambda C_v}e^{-\frac{16\pi^2k^2\sigma}{\lambda^2C_v}t}e^{-\frac{4\pi kz}{\lambda}}|_{z=0}    \label{eq7}
\end{equation}
In this equation $R$, $k$=0.79, $I_0$=2 $\mu$J/cm$^2$, $\lambda$=800 nm, $C_v$, and $\sigma$ are the reflectivity, dielectric extinction coefficient (see Supplementary Sec. III\cite{Sup}), pump fluence, pump wavelength, total specific heat, and thermal conductivity, respectively. The change in reflection can be calculated as: $\Delta R=R|_{T_0}^{T_0+\Delta T(t=0 ps)}$. The experimentally measured (blue dots) and smoothed (black line) $T$-dependent reflectivity $R$ are shown in the inset of Fig. \ref{SSlowDyn}(a). With the above equations and $A_3$, the 1D thermal transport model gives total specific heat $C_v$ of Ca$_3$Ru$_2$O$_7$ as shown by open circles in Fig. \ref{SSlowDyn}(a). The $C_v$ extracted by the 1D model qualitatively agrees well with calorimetry data \cite{Ke2014} in both the magnitude and $T$-dependence, confirming that the slow dynamics arise from thermal dissipation. The overall temperature change due to optical heating is determined to be smaller than 1 K above 30 K and increases to $\sim$7 K at 10 K, as shown in the inset of Fig. \ref{SSlowDyn}(b).

\begin{figure}[h]
 \includegraphics[width=0.78\columnwidth]{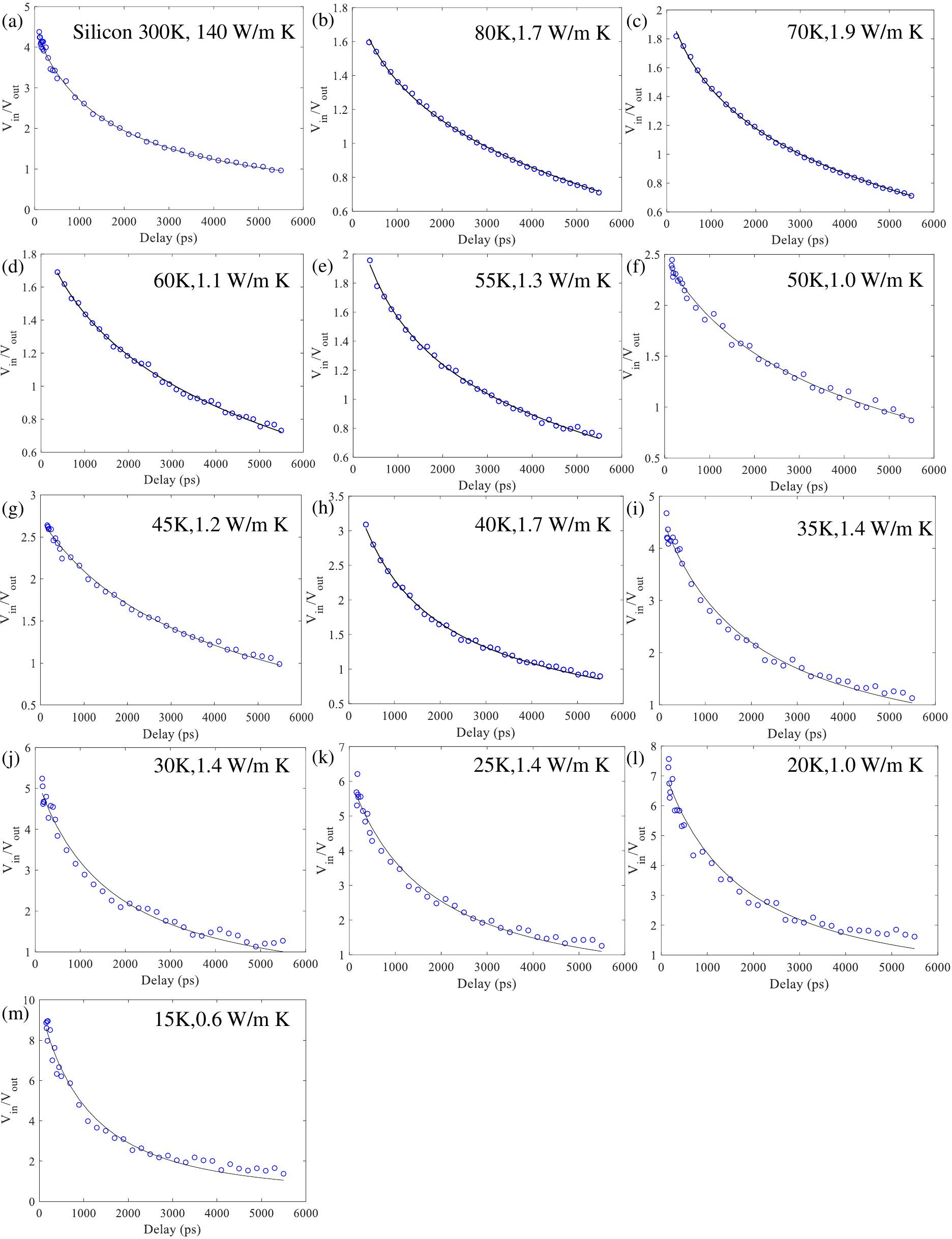}
 \caption{(a) Calibration measurement on silicon gives thermal conductivity of 140 W/m K, which lies within the expected range of 130-150 W/m K, confirming the validity of the TDTR setup. (b)-(m) $T$-dependent thermal conductivity measurements on Ca$_3$Ru$_2$O$_7$ gives thermal conducitivity of 0.6-1.9 W/m K in temperature range of 15-80 K.
\label{STDTR}}
\end{figure}

Similarly, from the time-dependent behavior of equation (\ref{eq7}), $C_v$, and relaxation time $\tau_{d3}$, the thermal conductivity of Ca$_3$Ru$_2$O$_7$ can be calculated as $\sigma=\frac{\lambda^2C_v}{16\pi^2k^2\tau_{d2}}$. This quantity is plotted in Fig. \ref{SSlowDyn}(b) using open circles. To compare with these results, Time-Domain Thermoreflectance (TDTR) measurements were carried out on aluminum coated Ca$_3$Ru$_2$O$_7$ as a function of temperature using 100 fs laser pulses at 800 nm with a repetition rate of 80 MHz.\cite{Schmidt2008,Schmidt20082} The pump was frequency doubled by a BiBO crystal and focused down to 20 $\mu$m of $1/e^2$ spot diameter, while probe at 800 nm had a spot size of 10 $\mu$m. The pump beam was modulated with an electro-optic modulator operating at 4.95 MHz. (For TDTR data, see Fig. \ref{STDTR}) The thermal conductivity measured on Ca$_3$Ru$_2$O$_7$ by the TDTR method is about 1.5 W/m K with very weak $T$-dependence between 10 K and 80 K (black line in Fig. \ref{SSlowDyn}(b)), which is of a similar order of magnitude as the values given by the 1D model. Note that this 1D model ignores the in-plane thermal transport and the possible anisotropy in thermal conductivity, hence a perfect match to TDTR results is not expected.

\section{III. Debye temperature and Dielectric constants of $Ca_3Ru_2O_7$}
The Debye temperature, $\theta_D$, of Ca$_3$Ru$_2$O$_7$ is calculated as follows. The lattice contribution to the specific heat in the low temperature limit obeys: 
\begin{equation}
C_v=\beta T^3, T\to0K \label{eqS3}
\end{equation}
with $\beta$=0.14 mJ/mol$_{Ru}$K$^4$.\cite{Yoshida2004} Under Debye's macroscopic picture of the phonon spectrum, we also have:
\begin{equation}
C_v=\frac{12\pi^4}{5}R_{gc}(\frac{T}{\theta_D})^3   , T\to0K \label{eqS4}
\end{equation}
where $R_{gc}$ is the gas constant. By comparing equations (\ref{eqS3}) and (\ref{eqS4}), we have $\theta_D\approx437$ K.

Dielectric constants of Ca$_3$Ru$_2$O$_7$ at room temperature were determined by spectroscopic ellipsometry measurements. Based on literature report on infrared spectroscopy study\cite{Lee2007}, there is no change in the dielectric constants at 800 nm between 10 K and $>T_C$. As an estimation, dielectric extinction coefficient at 800 nm is taken as $k$=0.79 over the entire temeprature range.
\begin{figure}[h]
 \includegraphics[width=0.75\columnwidth]{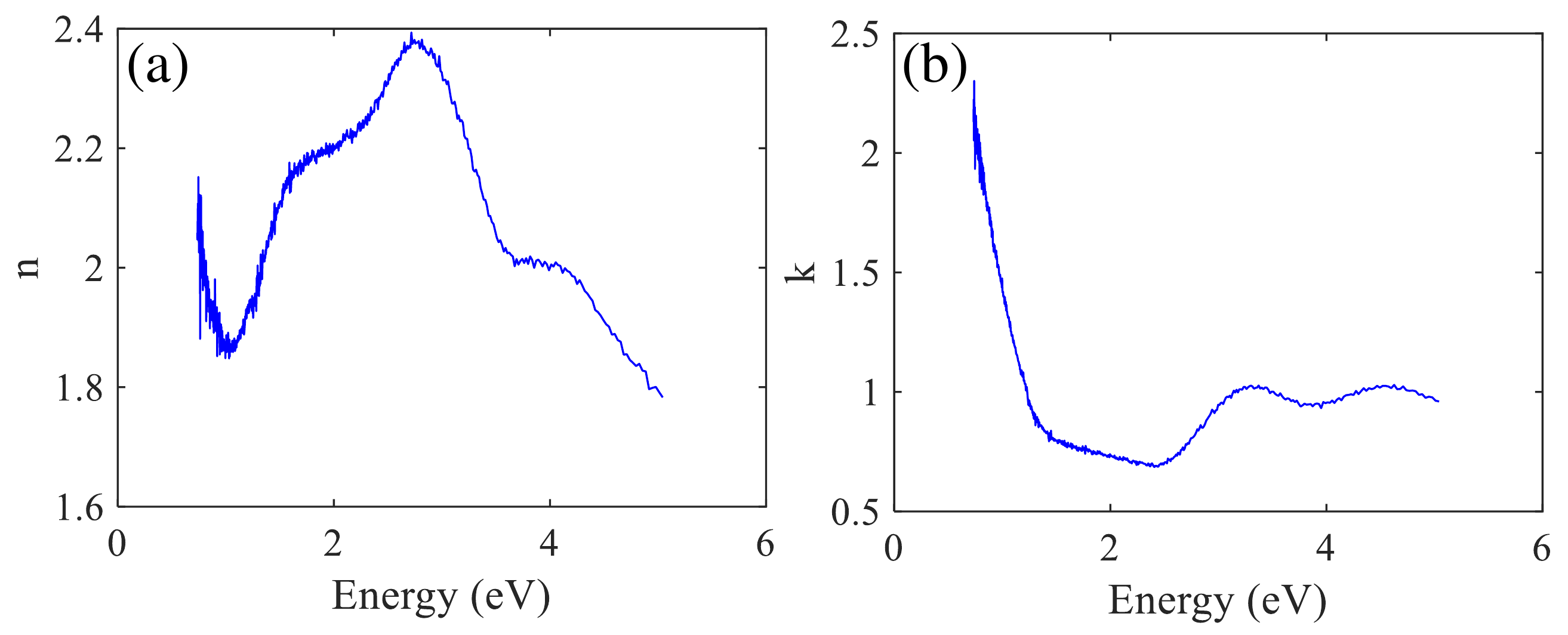}
 \caption{Dielectric constants, (a)$n$ and (b)$k$, of Ca$_3$Ru$_2$O$_7$ at room temperature. 
\label{Snk}}
\end{figure}

\section{IV. Suppression of electronic DOS in AFM-$b$ state revealed by DFT+$U$+$SOC$}

\begin{figure}[h]
 \includegraphics[width=1\columnwidth]{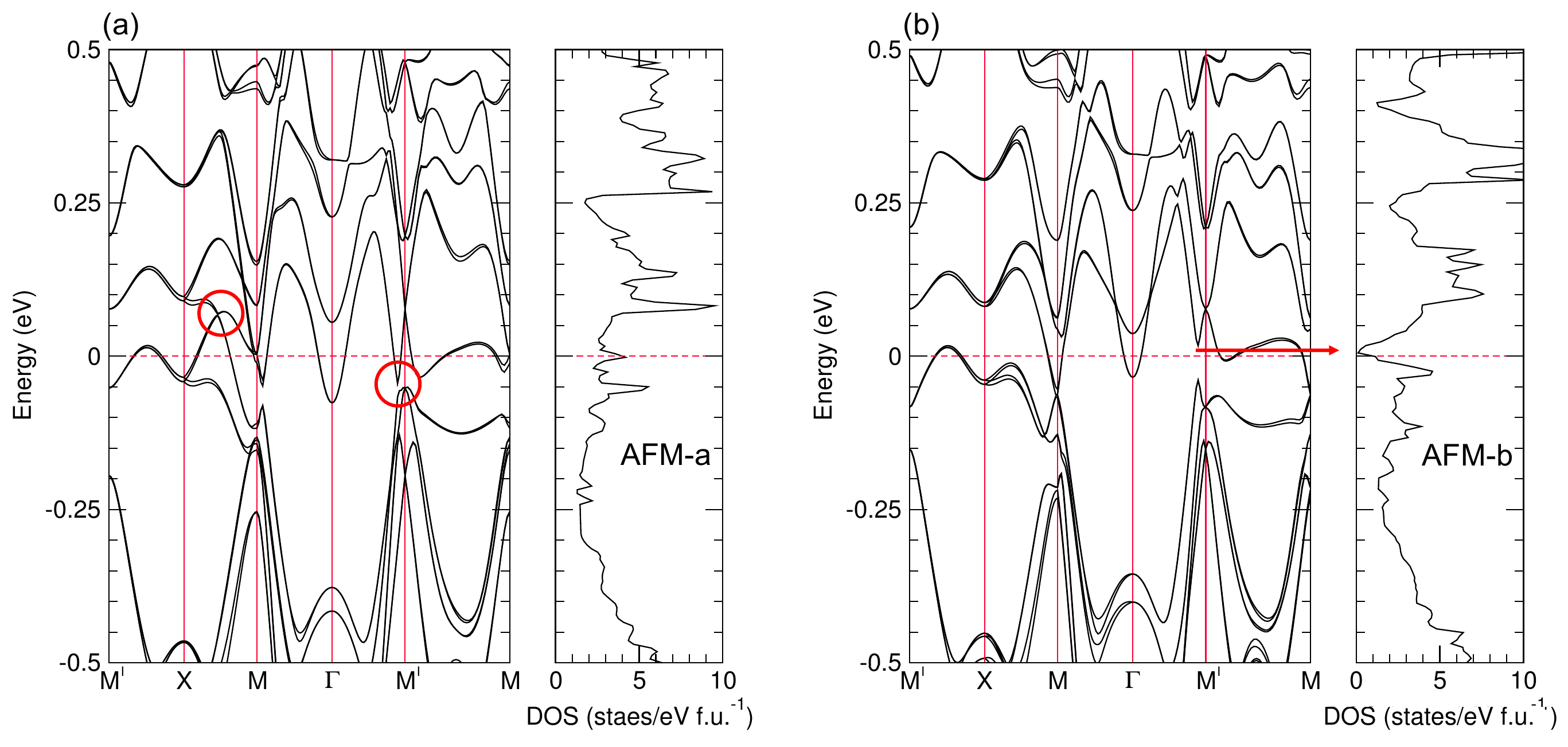}
 \caption{(a) Band structure and DOS in AFM-a state, which is the ground state of Ca$_3$Ru$_2$O$_7$ above $T_C$. (b) Band structure and DOS in AFM-b state, which is the ground state of Ca$_3$Ru$_2$O$_7$ below $T_C$. A suppression of DOS is observed compared to AFM-a state in DOS near E$_F$. The minimum point of this dip locates at $\sim7$ meV above E$_F$.
\label{SDFT}}
\end{figure}
To explore the gap related feature in electronic density-of-states in the AFM-b ground state below $T_C$. We performed electronic structure calculations based on density functional theory (DFT) within the revised Perdew-Burke-Ernzerh of exchange-correlation functional revised for solids\cite{Perdew2008,Kresse1996} as implemented in the Vienna Ab-initio Simulation Package\cite{Kresse1999} with the projector-augmented wave method\cite{Blochl1994} to treat the core and valence electrons using the following electronic configurations 3s2 3p6 4s2 (Ca), 5s2 4d6 (Ru), and 2s2 2p4 (O), and a 500 eV plane wave cutoff. Electron correlations in Ru-4d  electrons were treated using the Hubbard-$U$ ($U_{eff}$ = 1.2 eV) method within the Dudarev formalism\cite{MacDonald2000}.  Spin-polarized calculations with non-collinear AFM-b and AFM-a spin order were imposed on the Ru atoms. A 7 $\times$ 7 $\times$ 5 Monkhorst-Pack $k$-point mesh\cite{Pack1977} and Gaussian smearing (20 meV width) was used for the Brillouin zone (BZ) sampling and integrations.

We find that, the DOS in AFM-a state exhibits a peak feature near E$_F$, qualitatively agrees with our single Gaussian shape of DOS above $T_C$. It also should be noted that the DOS extracted from the fast relaxation dynamics under TTM only captures the dynamics of electrons near E$_F$ within the thermal activation energy scale $k_BT$, thus not sensitive to the electronic states far away from the E$_F$. For the AFM-b ground state below $T_C$, band structure and DOS by DFT calculations predict a dip feature near E$_F$ with a minimum position at $\sim$7 meV. This is consistent with the $T$-dependent DOS obtained using TTM, where a dip feature centered at $\sim$4.3 meV shows up as approaching 0 K. The position of this feature yielded by DFT and TTM analysis qualitatively agree well with each other.

\section{V. Failure of the Rothwarf-Taylor model for the fast dynamics $A_1$}
\begin{figure}[h]
 \includegraphics[width=0.75\columnwidth]{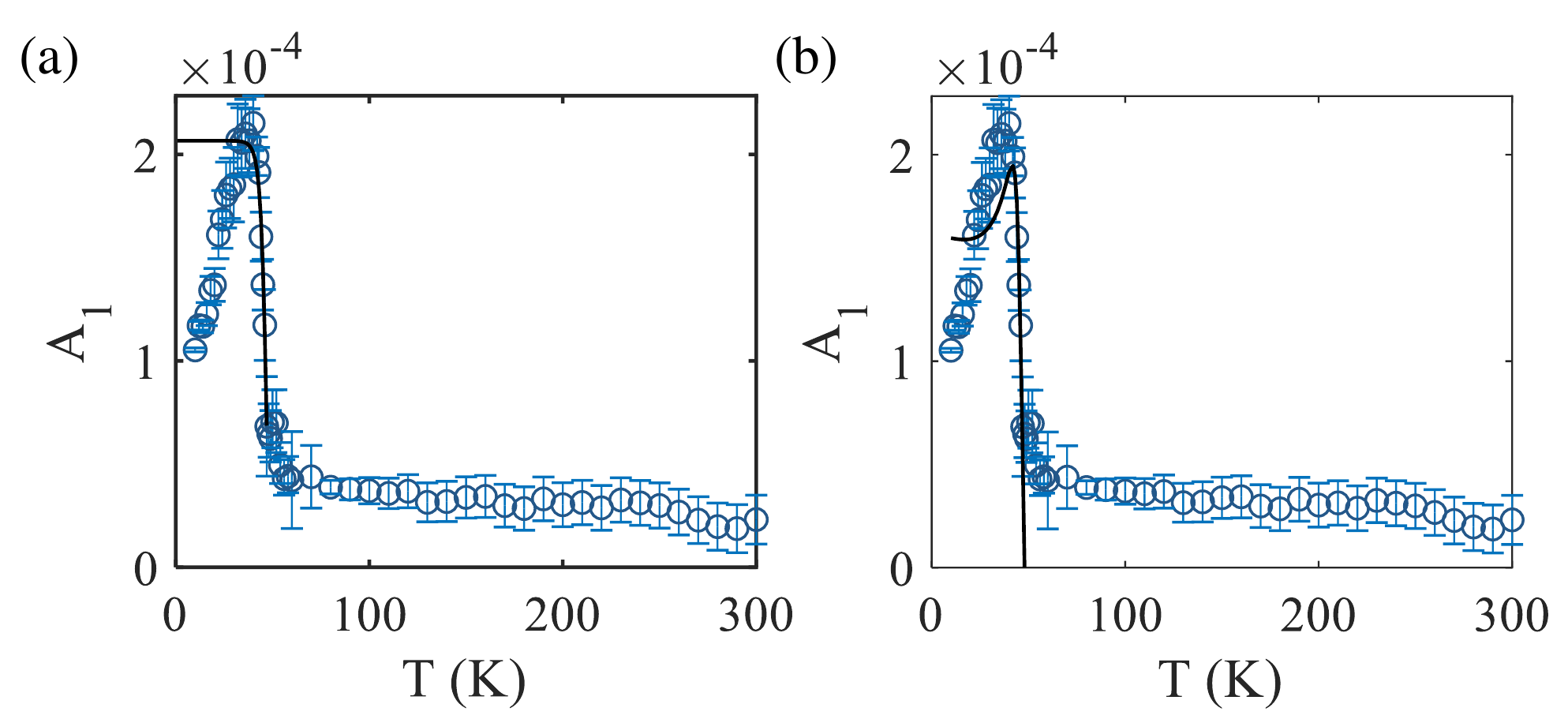}
 \caption{(a) R-T model (black line) fails to explain the amplitude of the fast dynamics $A_1$ (open circles). (b) Kabanov's model (equation (\ref{eqS5})) describing gap related dynamics fails to fit $A_1$. 
\label{SFastDynRTfit}}
\end{figure}
R-T analysis of $A_1$ fails to capture the behavior at low temperature ($T<$ 30 K), as shown in Fig. \ref{SFastDynRTfit}(a). Kabanov, et al. proposed another model to explain the gap related QP relaxation dynamics in YBa$_2$Cu$_3$O$_{7-\delta}$, as follow. \cite{Kabanov1999} The pump induced reflectivity $\Delta R/R$ is proportional to the number of photoexcited QPs, $n_{pe}$, which is given by:

\begin{equation}
n_{pe}\propto\frac{1/(\Delta(T)+k_BT/2)}{1+\alpha\sqrt{\frac{2k_BT}{\pi\Delta(T)}}exp(-\Delta(T)/k_BT)} \label{eqS5}
\end{equation}
where $\Delta(T)$ is the $T$-dependent gap size, and $\alpha$ is a constant. The fits of equation (\ref{eqS5}) to $A_1$ cannot explain the experimental data. (Fig. \ref{SFastDynRTfit}(b)) Hence, the dynamics related to $A_1$ shouldn't be related to the gap behavior in Ca$_3$Ru$_2$O$_7$. Instead, a $T$-dependent DOS combined with Two-Temperature Model (TTM) is used to explain this dynamics in greater detail, as discussed in the main text.

\section{VI. The Resistivity Crossover temperature $T^*$}

\begin{figure}[h]
 \includegraphics[width=1\columnwidth]{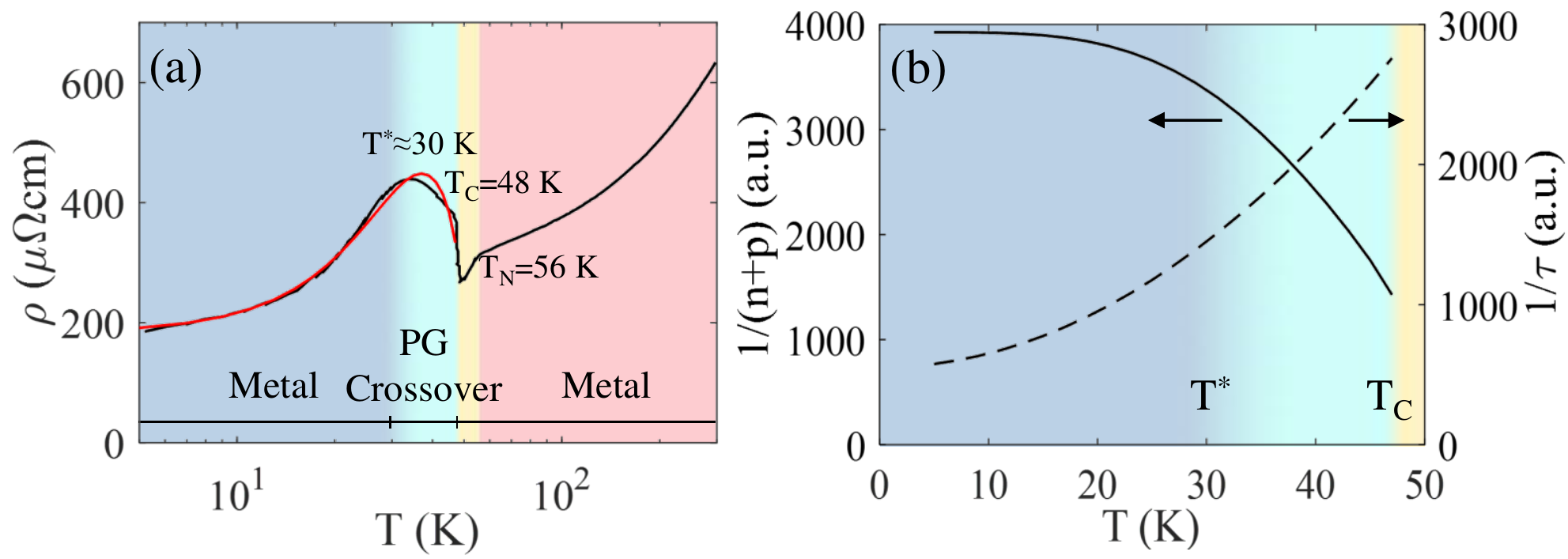}
 \caption{(a) In-plane resistivity of Ca$_3$Ru$_2$O$_7$ (black line) and its fit below $T_C$ (red line) reveal a pseudogap crossover region between $T^*$ and $T_C$. (b) $1/(n+p)$ and $1/\tau$ extracted from the fit.
 \label{SRhoFit}}
\end{figure}

\section{VII. The Rothwarf-Taylor model}
\begin{figure}[h]
 \includegraphics[width=0.5\columnwidth]{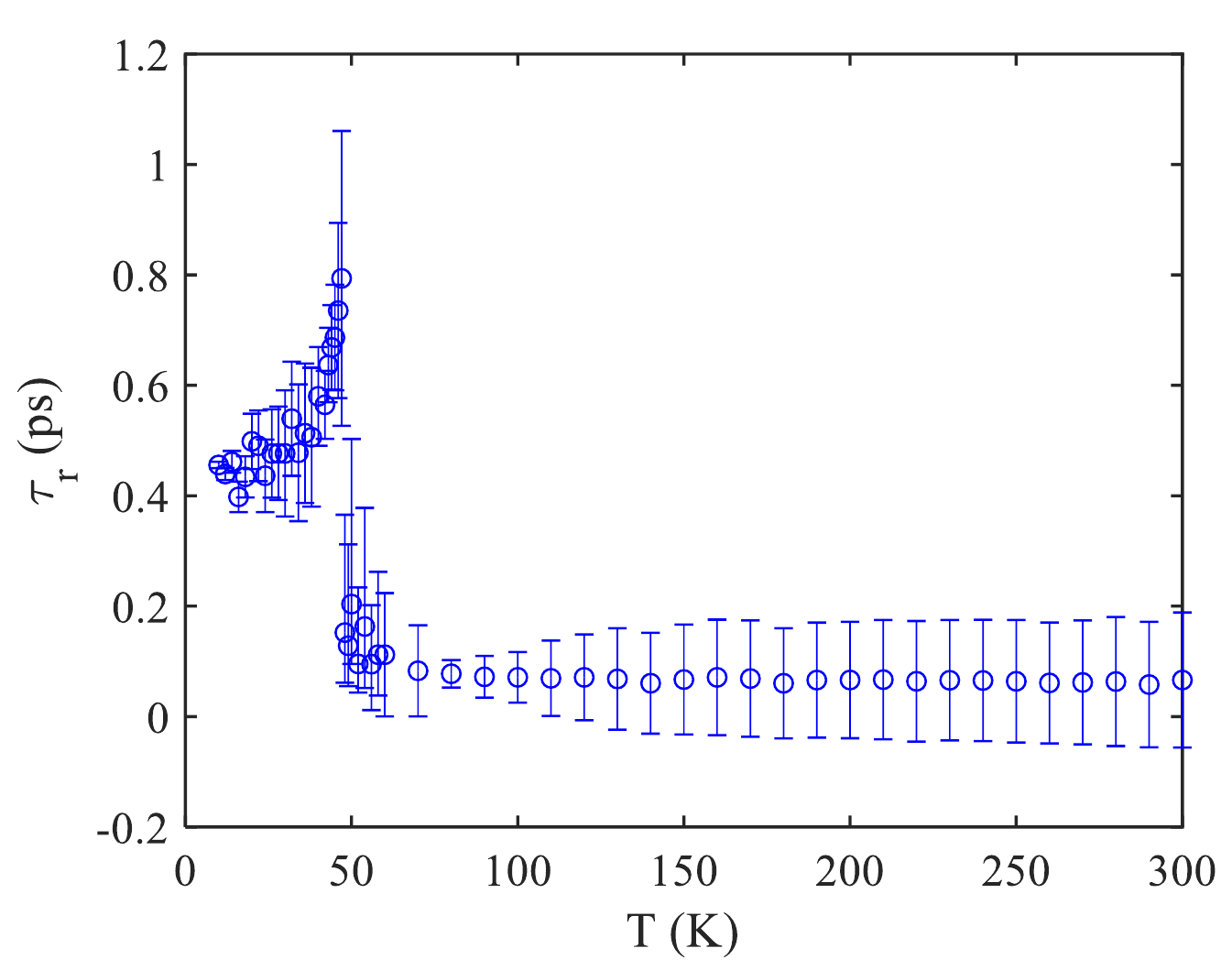}
 \caption{$T$-dependent rise time $\tau_r$ in Ca$_3$Ru$_2$O$_7$. The significant increase below $T_C$ signifies the strong phonon bottleneck regime.
 \label{Str}}
\end{figure}

The relaxation dynamics of coupled systems of quasiparticles and high energy phonons are often analyzed using the phenomenological Rothwarf-Taylor (R-T) model \cite{Rothwarf1967}. The key experimental parameters in the R-T model are the bare QP recombination rate $\gamma_{r}$, the rate of QP excitation by a HEP, $\gamma_{pc}$, and the escape rate of HEPs, $\gamma_{esc}$. Various regimes are realized depending on these rates \cite{Demsar2001, Gedik2004, Kabanov2005, Torchinsky2010}. Bimolecular recombination dynamics, with a relaxation rate that depends linearly on pump fluence ($F$), result if $\gamma_{r} \gg \gamma_{pc}$ or $\gamma_{esc} \gg \gamma_{pc}$, and the bare recombination rate of QPs, $\gamma_{pc}$, can be extracted. However, if $\gamma_{pc}$ is the fastest rate, the result is a strong phonon bottleneck with an $F$-independent relaxation rate $\gamma_{esc}$ determined by the escape or decay of HEPs. The observation of the significant slow rise time, $\tau_r$ (Fig. \ref{Str}), below $T_C$, as in Ca$_3$Ru$_2$O$_7$, strongly indicates that the system is in the strong phonon bottleneck regime \cite{Kabanov2005}. The observation of strongly temperature dependent slow rise dynamics in the strong bottleneck regime is typically explained by an increase in scattering between hot electrons and phonons in the presence of a gap in the electronic DOS at $E_F$. This results in an initial excess population of HEPs compared to the quasiequilibrium condition of the bottleneck, which can excite more electrons across the gap \cite{Demsar2003}.

The Rothwarf-Taylor model admits analytical solutions in various limiting regimes \cite{Kabanov2005}. In the small photo excitation limit, the amplitudes \(A\) and relaxation rates \(\tau^{-1}\) of the transient reflectivity signal are related to the number densities of photoexcited and thermally activated QPs, \(n_{S}\) and \(n_{T}\) respectively , as \(A\) \(\propto\) \(n_{S}-n_{T}\) and \(\tau^{-1}\) \(\propto\) \(n_{S}+n_{T}\). This statement may be expressed as:
\begin{equation}
n_T(T)\propto\frac{A(T\to0)}{A(T)}-1\label{eqS1}
\end{equation}
\begin{equation}
\tau^{-1}(T) = \Gamma[\delta+2n_{T}(T)](\Delta(T)+\alpha \Delta(T)T^{4})\label{eqS2}
\end{equation}
where $\Gamma$, $\delta$, and $\alpha$ are T-independent fitting parameters, $\Delta(T)+\alpha \Delta(T)T^{4}$ describes the gap size dependence of HEP decay rate, with an upper limit of $\alpha<52/(\theta^3_DT_{min})$, $\theta_D$ and $T_{min}$=10 K being Debye temperature and minimum temperature of the experiment.

\bibliography{Refs}

\end{document}